\documentclass[aps,pra,superscriptaddress,twocolumn,showpacs]{revtex4}
\usepackage{mathcomp}
\usepackage{amsmath}
\usepackage{amsfonts}
\usepackage{amssymb}
\usepackage{graphics}
\usepackage{epsfig}
\usepackage{amsmath}

\newtheorem{theorem}{Theorem}
\newtheorem{lemma}[theorem]{Lemma}
\newtheorem{corollary}[theorem]{Corollary}
\newtheorem{criterion}[theorem]{Criterion}

\begin{document}

\title{Correlation matrices of two-mode bosonic systems}
\author{Stefano Pirandola}
\affiliation{Research Laboratory of Electronics, MIT, Cambridge, Massachusetts 02139, USA}
\author{Alessio Serafini}
\affiliation{Department of Physics \& Astronomy, University College London, Gower Street,
London WC1E 6BT, United Kingdom}
\author{Seth Lloyd}
\affiliation{Research Laboratory of Electronics, MIT, Cambridge, Massachusetts 02139, USA}
\affiliation{Department of Mechanical Engineering, MIT, Cambridge, Massachusetts 02139,
USA}

\begin{abstract}
We present a detailed analysis of all the algebraic conditions an arbitrary $%
4\times 4$ symmetric matrix must satisfy in order to represent the
correlation matrix of a two-mode bosonic system. Then, we completely clarify
when this arbitrary matrix can represent the correlation matrix of a
separable or entangled Gaussian state. In this analysis, we introduce new
and alternative sets of conditions, which are expressed in terms of local
symplectic invariants.
\end{abstract}

\maketitle

\section{Introduction}

Statistical moments of second order represent a key element of the quantum
mechanical paradigm. Besides providing the `language' in which the
uncertainty principles are expressed, they serve as indicators in a number
of applications of the theory, with both applied and fundamental interest.
In particular, second moments are central to the description of bosonic
fields in second quantization (as is the case, for instance, in quantum
optics, where second order coherence is characterized in terms of second
moments) and of non-relativistic particles in first quantization. Such
systems do, in fact, share the same formal description \cite{numdegNote},
which hence extends its domain to a variety of fields ranging from atomic
physics to quantum optics, from superconductors' physics to nano-mechanical
systems.

During the last decade, the rise of quantum information science has renewed
the focus on these areas because of their potential for coherent quantum
manipulations, and has concomitantly brought new problems and questions to
the attention of theorists, which resulted in the birth of the field of
\textquotedblleft continuous variable\textquotedblright\ (CV)\ quantum
information \cite{BrauRev} (see Refs.~\cite{CVQcomp,CVQKD,CVTelepo} for some
literature on CV quantum computation, CV quantum teleportation, and CV
quantum key distribution, respectively). A systematic analysis of the
properties of continuous variable quantum states inferred from the structure
of their second moments has been thereby carried out, which lead to a
well-established, extensive theoretical picture (see, for instance, \cite%
{Simon,Duan,wernerwolf,AlexHeis}). Such an analysis proved to be most
relevant also in view of the experimental prominence of the class of \emph{%
Gaussian states} \cite{EisertGauss,NapoliGauss}, which are completely
determined by their first and second moments. In first, seminal endeavours
\cite{Simon,Duan}, the qualitative characterization of the quantum
correlations (\textquotedblleft entanglement\textquotedblright ) of Gaussian
states of two degrees of freedom has been successfully achieved. Now, while
very well established and relatively simple, this result is still often
expressed in a non-rigourous or incomplete manner, which is prone to
confound the unacquainted reader. Because it constitutes one of the basic
building blocks on which the theoretical characterization of Gaussian states
has been constructed, it seems to us extremely important for it to be
re-derived and re-expressed in a rigourous manner and full detail: this is
one of the motifs and central aims of the present paper.

In general, the main question we will address and rigourously answer is the
following:

\begin{description}
\item[(i)] \emph{What are the algebraic conditions that a }$4\times 4$\emph{%
\ real symmetric matrix }$\mathbf{V}$\emph{\ must satisfy in order to
represent the correlation matrix of a two-mode bosonic system?} \cite%
{cm_note}
\end{description}

\noindent Then, we shall move on to answer a closely connected question:

\begin{description}
\item[(ii)] \emph{What are the algebraic conditions to be satisfied by }$%
\mathbf{V}$\emph{\ in order to represent the correlation matrix of a
separable (or entangled) Gaussian state of two bosonic modes?}
\end{description}

\noindent Both these questions are thoroughly answered providing complete
sets of conditions, which are expressed in terms of global or local
symplectic invariants. In particular, the set of local conditions, i.e.,
given in terms of local invariants, is completely new in literature. Then,
by specifying some of these algebraic results in the case of
positive-definite matrices, we can make a direct comparison with the
previous work of Ref.~\cite{Simon} and provide a rigorous and correct
interpretation of its seminal results.

The paper is organized as follows. In Sec.~\ref{SecGEN} we review basic
notions about bosonic states, correlation matrices and symplectic
transformations. In Sec.~\ref{ToolsSEC} we present the mathematical tools to
be used in the derivations of Secs.~\ref{SECgenuine}\ and~\ref%
{SECseparability}. In Sec.~\ref{SECgenuine} we provide two sets of algebraic
conditions for the physical genuinity of the correlation matrix of two
bosonic modes. These conditions are expressed in terms of global or local
symplectic invariants. In Sec.~\ref{SECseparability} we provide similar
conditions for separability. Next, in Section~\ref{Comparison}, we specify
some of our results for making a direct comparison with the previous
achievements of Ref.~\cite{Simon}. Finally, Sec.~\ref{SECsummary} is for
conclusions.

Notice that in the paper we will denote by $\mathcal{M}(n,\mathbb{R})$ the
set of $n\times n$ real matrices. Then, we use the compact notation
\begin{equation}
\mathcal{S}(n,\mathbb{R})=\{\mathbf{M}\in \mathcal{M}(n,\mathbb{R}):\mathbf{%
M=M}^{T}\}~,
\end{equation}%
for the set of the $n\times n$ symmetric real matrices, and
\begin{equation}
\mathcal{P}(n,\mathbb{R})=\{\mathbf{M}\in \mathcal{S}(n,\mathbb{R}):\mathbf{M%
}>0\}~,
\end{equation}%
for the set of the $n\times n$ positive-definite real matrices. We will also
consider the set (group) of proper rotations%
\begin{equation}
\mathcal{SO}(n)=\{\mathbf{M}\in \mathcal{M}(n,\mathbb{R}):\mathbf{M}^{T}%
\mathbf{M=I},~\det \mathbf{M}=1\}~,
\end{equation}%
where $\mathbf{I}$\ is the identity matrix.

\section{Bosonic systems and symplectic transformations\label{SecGEN}}

\subsection{Correlation matrix of a bosonic system}

Let us consider a bosonic system composed by $n$ modes, labeled by an index $%
k$. Such a system can be described by an infinite-dimensional Hilbert space $%
\mathcal{H}=\otimes _{k=1}^{n}\mathcal{H}_{k}$ and a vector of quadrature
operators $\mathbf{\hat{x}}^{T}:=(\hat{q}_{1},\hat{p}_{1},\cdots ,\hat{q}%
_{n},\hat{p}_{n})$. In particular, these operators satisfy the commutation
relations
\begin{equation}
\lbrack \hat{x}_{l},\hat{x}_{m}]=2i\mathbf{\Omega }_{lm}~,  \label{CCReq}
\end{equation}%
where $l,m=1,\cdots ,2n$, and $\mathbf{\Omega }_{lm}$ are the entries of the
simplectic form%
\begin{equation}
\mathbf{\Omega }=\bigoplus\limits_{k=1}^{n}\boldsymbol{\omega }~,~%
\boldsymbol{\omega }:=\left(
\begin{array}{cc}
0 & 1 \\
-1 & 0%
\end{array}%
\right) ~.  \label{Symplectic_Form}
\end{equation}%
An arbitrary state of the bosonic system is identified with a density
operator $\rho $ acting on the Hilbert space $\mathcal{H}$ [we denote by $%
\mathcal{D}(\mathcal{H})$ the space of density operators acting on $\mathcal{%
H}$]. An arbitrary density operator $\rho \in \mathcal{D}(\mathcal{H})$ has
an equivalent representation in a real symplectic space $\mathcal{K}=%
\mathcal{K}(\mathbb{R}^{2n},\mathbf{\Omega })$ called the \emph{phase space}%
. This is a real vector space which is spanned by the singular eigenvalues $%
\mathbf{x}^{T}\mathbf{=}(q_{1},p_{1},\cdots ,q_{n},p_{n})$ of $\mathbf{\hat{x%
}}^{T}$ (representing the \textquotedblleft continuous
variables\textquotedblright\ of the system) and associated to a symplectic
product $\mathbf{u}\cdot \mathbf{v}=\mathbf{u}^{T}\mathbf{\Omega v}$. In
this space, a quantum state is fully described by a quasi-probability
distribution known as the Wigner function $W=W(\mathbf{x})$. In general,
such a function is fully characterized by the entire set of its statistical
moments \cite{QObook}. However, in the particular case of Gaussian states,
the Wigner function is Gaussian and, therefore, fully characterized by the
first and second moments only. These two moments are also known as the
\textit{displacement vector} $\mathbf{d}:=\left\langle \mathbf{\hat{x}}%
\right\rangle $ and the \textit{correlation matrix} (CM) $\mathbf{V}$, whose
generic entry is defined by%
\begin{equation}
\mathbf{V}_{lm}:=\frac{1}{2}\langle \Delta \hat{x}_{l}\Delta \hat{x}%
_{m}+\Delta \hat{x}_{m}\Delta \hat{x}_{l}\rangle  \label{CM_definition}
\end{equation}%
where $\Delta \hat{x}_{l}:=\hat{x}_{l}-\langle \hat{x}_{l}\rangle $.
According to the definition of Eq.~(\ref{CM_definition}), the CM of $n$
bosonic modes is a real and symmetric matrix in $2n$ dimension, i.e., $%
\mathbf{V}\in \mathcal{S}(2n,\mathbb{R})$. As a direct consequence of Eq.~(%
\ref{CCReq}), such a matrix must also satisfy the uncertainty principle \cite%
{Simon3,AlexHeis}
\begin{equation}
\mathbf{V}+i\mathbf{\Omega }\geq 0~.  \label{BONA_FIDE}
\end{equation}%
In other words, an arbitrary $\mathbf{V}\in \mathcal{S}(2n,\mathbb{R})$ is a
\emph{bona fide} quantum CM if and only if Eq.~(\ref{BONA_FIDE}) holds.
Equivalently, we can introduce the set of $n$-mode quantum CM's to be
defined as%
\begin{equation}
q\mathcal{CM}(n):=\left\{ \mathbf{V}\in \mathcal{S}(2n,\mathbb{R}):\mathbf{V}%
+i\mathbf{\Omega }\geq 0\right\} ~.
\end{equation}%
Notice that the condition of Eq.~(\ref{BONA_FIDE}) implies a first relevant
constraint on the matrix $\mathbf{V}$:

\begin{lemma}[Definite positivity of $\mathbf{V}$]
\label{LemmaPOS}For every $\mathbf{V}\in \mathcal{S}(2n,\mathbb{R})$
satisfying $\mathbf{V}+i\mathbf{\Omega }\geq 0$, one has
\begin{equation}
\mathbf{V}>0~.
\end{equation}
\end{lemma}

\bigskip

\noindent \textbf{Proof.~} Let $\mathbf{u}\in \mathbb{R}^{2n}$, then $0\leq
\mathbf{u}^{T}\mathbf{Vu}+i\mathbf{u}^{T}\mathbf{\Omega u}=\mathbf{u}^{T}%
\mathbf{Vu}$ because $\mathbf{\Omega }$ is anti-symmetric. Hence $\mathbf{V}%
\geq 0$. To prove definite positivity suppose, \emph{ad absurdum}, that a
non-trivial real vector $\mathbf{u}_{0}$ exists such that $\mathbf{u}_{0}^{T}%
\mathbf{Vu}_{0}=0$. Another vector $\mathbf{u}_{1}\in \mathbb{R}^{2n}$ such
that $\mathbf{u}_{1}^{T}\mathbf{\Omega u}_{0}\neq 0$ always exists as $%
\mathrm{nul}(\mathbf{\Omega })=0$. As a consequence, one can always
construct a set of complex vectors $\mathbf{z}=\mathbf{u}_{0}+ia\mathbf{u}%
_{1}$, for $a\in \mathbb{R}$, such that
\begin{equation}
0\leq (\mathbf{z}^{\ast })^{T}(\mathbf{V}+i\mathbf{\Omega })\mathbf{z}=2a%
\mathbf{u}_{1}^{T}\mathbf{\Omega u}_{0}+a^{2}\mathbf{u}_{1}^{T}\mathbf{Vu}%
_{1}~.
\end{equation}%
Values of $a$ such that the inequality above is violated can always be found
regardless of the values of $\mathbf{u}_{1}^{T}\mathbf{\Omega u}_{0}$ and $%
\mathbf{u}_{1}^{T}\mathbf{Vu}_{1}$. This implies $\mathbf{u}^{T}\mathbf{Vu}%
\neq 0$ for every $\mathbf{u}\in \mathbb{R}^{2n}$ and, therefore, $\mathbf{V}%
>0$.$~\blacksquare $

According to Lemma~\ref{LemmaPOS}, we then have%
\begin{equation}
q\mathcal{CM}(n)\subseteq \mathcal{P}(2n,\mathbb{R})~.
\end{equation}%
Furthermore, it is trivial to show positive-definite matrices which violate
Eq.~(\ref{BONA_FIDE}), so that we actually have%
\begin{equation}
q\mathcal{CM}(n)\subset \mathcal{P}(2n,\mathbb{R})~.
\end{equation}%
Notice that definite positivity is the only requirement for a real symmetric
matrix to be a \emph{classical} correlation matrix.

\subsection{Symplectic transformations}

The most general real linear transformation of the quadratures
\begin{equation}
\mathbf{S}:\mathbf{\hat{x}}\longrightarrow \mathbf{\hat{x}}^{\prime }:=%
\mathbf{S\hat{x}}~,  \label{Linear_map}
\end{equation}%
must preserve Eq.~(\ref{CCReq}) in order to be a physical operation. This
happens when the matrix $\mathbf{S}\in \mathcal{M}(2n,\mathbb{R})$ preserves
the symplectic form of Eq.~(\ref{Symplectic_Form}), i.e.,%
\begin{equation}
\mathbf{S\Omega S}^{T}=\mathbf{\Omega }~.  \label{Sympl_cond}
\end{equation}%
The set of all the matrices $\mathbf{S}\in \mathcal{M}(2n,\mathbb{R})$
satisfying Eq.~(\ref{Sympl_cond}) forms the so-called \emph{real symplectic
group}
\begin{equation}
\mathcal{S}_{p}(2n,\mathbb{R}):=\{\mathbf{S}\in \mathcal{M}(2n,\mathbb{R}):%
\mathbf{S\Omega S}^{T}=\mathbf{\Omega }\}~,
\end{equation}%
whose elements are called \emph{symplectic} or \emph{canonical
transformations}. As a consequence, the most general real linear
transformation in phase space $\mathbf{S}:\mathbf{x}\longrightarrow \mathbf{x%
}^{\prime }:=\mathbf{Sx}$ must be symplectic. Its action on the Wigner
function is simply given by $W(\mathbf{x})\longrightarrow W(\mathbf{S}^{-1}%
\mathbf{x})$, so that the displacement is linearly modified while the CM is
transformed according to the congruence%
\begin{equation}
\mathbf{V}\longrightarrow \mathbf{SVS}^{T}~.  \label{CM_congruence}
\end{equation}%
Symplectic transformations are very important since every $\mathbf{S}$
acting in the phase space $\mathcal{K}$ corresponds to a \textit{Gaussian}
unitary $\hat{U}(\mathbf{S})$ acting on the Hilbert space $\mathcal{H}$,
i.e., a unitary operator preserving the Gaussian statistics of the quantum
states. These unitaries are the ones generated by bilinear Hamiltonians and
can always be decomposed into single-mode squeezers and multi-mode
interferometers \cite{BrauMessiah,EisertGauss}. In particular, local
symplectic transformations
\begin{equation}
\mathbf{S=}\bigoplus\limits_{k=1}^{n}\mathbf{S}_{k}\in \mathcal{S}_{p}(2,%
\mathbb{R})\oplus \cdots \oplus \mathcal{S}_{p}(2,\mathbb{R})
\label{Local_Symple}
\end{equation}%
correspond to local Gaussian unitaries
\begin{equation}
\hat{U}(\mathbf{S})=\bigotimes\limits_{k=1}^{n}\hat{U}_{k}~.
\end{equation}%
Local symplectic transformations can always be decomposed as products of
local rotations and local squeezings. In fact, thanks to the following
characterization%
\begin{equation}
\mathcal{S}_{p}(2,\mathbb{R})=\{\mathbf{S}\in \mathcal{M}(2,\mathbb{R}):\det
\mathbf{S}=1\}~,  \label{Sympl_Det_Equiv}
\end{equation}%
we have that every $\mathbf{S}\in \mathcal{S}_{p}(2,\mathbb{R})$ can be
expressed as a product of proper rotations%
\begin{equation}
\mathbf{R}(\varphi ):=\left(
\begin{array}{cc}
\sin \varphi & -\cos \varphi \\
\cos \varphi & \sin \varphi%
\end{array}%
\right) ~,
\end{equation}%
and squeezing matrices
\begin{equation}
\mathbf{S}(\xi ):=\left(
\begin{array}{cc}
\xi ^{1/2} & 0 \\
0 & \xi ^{-1/2}%
\end{array}%
\right) ~,~\xi >0~.  \label{Squez_matrix}
\end{equation}

Besides, it is also important to identify which quantities of a CM are
preserved under the application of symplectic transformations. In general,
for a given CM $\mathbf{V}$, we say that a functional
\begin{equation}
f:\mathbf{V}\rightarrow f(\mathbf{V})\in \mathbb{R}
\end{equation}%
is a (global) symplectic invariant if
\begin{equation}
f(\mathbf{V})=f(\mathbf{SVS}^{T})~,  \label{Sympl_inv}
\end{equation}%
for every $\mathbf{S}\in \mathcal{S}_{p}(2n,\mathbb{R})$. Then, we say that $%
f(\mathbf{V})$ is a local symplectic invariant if Eq.~(\ref{Sympl_inv})
holds for every $\mathbf{S}\in \mathcal{S}_{p}(2,\mathbb{R})\oplus \cdots
\oplus \mathcal{S}_{p}(2,\mathbb{R})$ \cite{global}. Notice that we can
extend the notion of symplectic invariance also to a property of a matrix.
For instance, the definite positivity of $\mathbf{V}$ is a global symplectic
invariant since $\mathbf{V}>0\Longrightarrow \mathbf{SVS}^{T}>0$.

\section{Symplectic analysis\label{ToolsSEC}}

Here, we review some basic tools that can be used for the symplectic
manipulation of the CM's. In particular, the central tool in this trade is
Williamson's theorem \cite{Williamson}, which ensures the possibility of
carrying out the symplectic diagonalization of real matrices in even
dimension \emph{under the definite positivity constraint} (as in the case of
the CM's).

\begin{lemma}[Williamson's theorem]
For every $\mathbf{V}\in \mathcal{P}(2n,\mathbb{R})$, there exists a
symplectic matrix $\mathbf{S}\in \mathcal{S}_{p}(2n,\mathbb{R})$ such that%
\begin{equation}
\mathbf{SV\mathbf{S}}^{T}=\left(
\begin{array}{ccccc}
\nu _{1} &  &  &  &  \\
& \nu _{1} &  &  &  \\
&  & \ddots &  &  \\
&  &  & \nu _{n} &  \\
&  &  &  & \nu _{n}%
\end{array}%
\right) :=\mathbf{W}>0~,  \label{Diag_Sympl2}
\end{equation}%
where the $n$ positive quantities $\{\nu _{1},\cdots ,\nu _{n}\}$ are called
the \textquotedblleft symplectic eigenvalues\textquotedblright\ of $\mathbf{V%
}$, and the diagonal matrix $\mathbf{W}$ is called the \textquotedblleft
Williamson form\textquotedblright\ (or \textquotedblleft normal
form\textquotedblright ) of $\mathbf{V}$.
\end{lemma}

The symplectic spectrum $\{\nu _{1},\cdots ,\nu _{n}\}$ can be computed as
the standard eigenspectrum of the matrix $|i\mathbf{\Omega V}|$ where the
modulus must be understood in the operatorial sense \cite{Spectrum}. The
corresponding Williamson form $\mathbf{W}$ is unique up to a permutation of
the symplectic spectrum (i.e., of the bosonic modes). Fixing this
permutation, the diagonalizing symplectic matrix $\mathbf{S}$ of Eq.~(\ref%
{Diag_Sympl2}) is defined up to local rotations $\oplus _{k=1}^{n}\mathbf{R}%
_{k}$ with $\mathbf{R}_{k}\in \mathcal{SO}(2)$. For the sake of
completeness, we report in Appendix~\ref{WILLIAMSONTHEO} a simple proof of
Williamson's theorem, originally presented in Ref.~\cite{SimonWILL} (see
also Ref.~\cite{NapoliGauss}). Using this proof, we show in Appendix~\ref%
{FINDING} an algorithm which finds the diagonalizing symplectic matrix $%
\mathbf{S}$ of Eq.~(\ref{Diag_Sympl2}). This algorithm is not the fastest
but it can be helpful in studying problems like the optimal discrimination
of Gaussian states \cite{Discrimination} and the Quantum Illumination \cite%
{Qillumination}.

Let us consider the case of a $4\times 4$ positive-definite matrix $\mathbf{V%
}\in \mathcal{P}(4,\mathbb{R})$, as in the case of CM's describing two
bosonic modes. This matrix can be expressed in the blockform
\begin{equation}
\mathbf{V}=\left(
\begin{array}{cc}
\mathbf{A} & \mathbf{C} \\
\mathbf{C}^{T} & \mathbf{B}%
\end{array}%
\right) ~,  \label{CM_2modes}
\end{equation}%
where $\mathbf{A},\mathbf{B}\in \mathcal{S}(2,\mathbb{R})$ and $\mathbf{C}%
\in \mathcal{M}(2,\mathbb{R})$. In this case the symplectic spectrum $\{\nu
_{1},\nu _{2}\}:=\{\nu _{-},\nu _{+}\}$ can be computed via the simple
formula \cite{Alex2modi}%
\begin{equation}
\nu _{\pm }=\sqrt{\frac{\Delta (\mathbf{V})\pm \sqrt{\Delta (\mathbf{V}%
)^{2}-4\det \mathbf{V}}}{2}}~,  \label{symple}
\end{equation}%
where%
\begin{equation}
\Delta (\mathbf{V}):=\det \mathbf{A}+\det \mathbf{B}+2\det \mathbf{C~.}
\label{seraliano2modes}
\end{equation}%
Here, the quantities $\det \mathbf{A}$, $\det \mathbf{B}$ and $\det \mathbf{C%
}$ are local symplectic invariants, while $\det \mathbf{V}$ and $\Delta (%
\mathbf{V})$ are global symplectic invariants, which can also be written as%
\begin{equation}
\det \mathbf{V}=\nu _{-}^{2}\nu _{+}^{2}~,~\Delta (\mathbf{V})=\nu
_{-}^{2}+\nu _{+}^{2}~.  \label{Global_Inv}
\end{equation}

Another important tool in the symplectic analysis is the reduction to \emph{%
standard form} by local symplectic transformations \cite{Simon,Duan}. In
general, such a reduction holds for symmetric matrices $\mathbf{V}\in
\mathcal{S}(4,\mathbb{R})$ with positive diagonal blocks, as we easily show
in the following. In particular, it can be applied to positive-definite
matrices $\mathbf{V}\in \mathcal{P}(4,\mathbb{R})$ and, therefore, to CMs $%
\mathbf{V}\in q\mathcal{CM}(2)$.

\begin{lemma}[Standard Form]
For every%
\begin{equation}
\mathbf{V}=\left(
\begin{array}{cc}
\mathbf{A} & \mathbf{C} \\
\mathbf{C}^{T} & \mathbf{B}%
\end{array}%
\right) \in \mathcal{S}(4,\mathbb{R})~,
\end{equation}%
with $\mathbf{A},\mathbf{B}>0$, there exists some $\mathbf{S}\in \mathcal{S}%
_{p}(2,\mathbb{R})\oplus \mathcal{S}_{p}(2,\mathbb{R})$ such that%
\begin{equation}
\mathbf{SVS}^{T}=\left(
\begin{array}{cccc}
a &  & c_{+} &  \\
& a &  & c_{-} \\
c_{+} &  & b &  \\
& c_{-} &  & b%
\end{array}%
\right) :=\mathbf{V}^{I}~,  \label{normal_FORM}
\end{equation}%
where the real parameters $a,b,c_{+},c_{-}$ satisfy%
\begin{equation}
\det \mathbf{A}=a^{2}~,~\det \mathbf{B}=b^{2}~,~\det \mathbf{C}=c_{+}c_{-}~,
\label{local_invariants}
\end{equation}%
and
\begin{equation}
\det \mathbf{V}=\det \mathbf{V}^{I}=(ab-c_{+}^{2})(ab-c_{-}^{2})~.
\label{Det_invariant}
\end{equation}
\end{lemma}

\bigskip

\noindent \textbf{Proof.~}Let us consider a pair of single-mode symplectic
transformations $\mathbf{S}_{A},\mathbf{S}_{B}\in \mathcal{S}_{p}(2,\mathbb{R%
})$ and a pair of single-mode proper rotations $\mathbf{R}(\theta _{A}),%
\mathbf{R}(\theta _{B})\in \mathcal{SO}(2)$. By applying the local
symplectic transformation%
\begin{equation}
\mathbf{S}=\mathbf{R}(\theta _{A})\mathbf{S}_{A}\oplus \mathbf{R}(\theta
_{B})\mathbf{S}_{B}  \label{Local_4proof}
\end{equation}%
to the matrix $\mathbf{V}$, we get%
\begin{equation}
\mathbf{SVS}^{T}=\left(
\begin{array}{cc}
\mathbf{A}^{\prime } & \mathbf{C}^{\prime } \\
\mathbf{C}^{\prime T} & \mathbf{B}^{\prime }%
\end{array}%
\right) ~,  \label{SVS}
\end{equation}%
where%
\begin{eqnarray}
\mathbf{A}^{\prime } &:&=\mathbf{R}(\theta _{A})~\left( \mathbf{S}_{A}~%
\mathbf{A~S}_{A}^{T}\right) ~\mathbf{R}(\theta _{A})^{T}~, \\
\mathbf{B}^{\prime } &:&=\mathbf{R}(\theta _{B})~\left( \mathbf{S}_{B}~%
\mathbf{B~S}_{B}^{T}\right) ~\mathbf{R}(\theta _{B})^{T}~, \\
\mathbf{C}^{\prime } &:&=\mathbf{R}(\theta _{A})~\left( \mathbf{S}_{A}~%
\mathbf{C~S}_{B}^{T}\right) \mathbf{~R}(\theta _{B})^{T}~.  \label{C_Block}
\end{eqnarray}%
Since $\mathbf{A,B}\in \mathcal{P}(2,\mathbb{R})$, we can apply Williamson's
theorem. This means that we can choose $\mathbf{S}_{A}$ and $\mathbf{S}_{B}$
such that%
\begin{eqnarray}
\mathbf{A}^{\prime } &=&\mathbf{R}(\theta _{A})~a\mathbf{I~R}(\theta
_{A})^{T}=a\mathbf{I~,} \\
\mathbf{B}^{\prime } &=&\mathbf{R}(\theta _{B})~b\mathbf{I~R}(\theta
_{B})^{T}=b\mathbf{I~,}
\end{eqnarray}%
where $a$ ($b$) is the symplectic eigenvalue of $\mathbf{A}$ ($\mathbf{B}$)
while the angle $\theta _{A}$ ($\theta _{B}$) is arbitrary. Since the pair $%
\{\theta _{A},\theta _{B}\}$ can be chosen freely, we can always choose a
pair $\{\bar{\theta}_{A},\bar{\theta}_{B}\}$ in Eq.~(\ref{C_Block}) such
that $\mathbf{C}^{\prime }=\mathrm{diag}(c_{+},c_{-})$ \cite{Algebra1}. As a
consequence, Eq.~(\ref{SVS}) is globally equal to Eq.~(\ref{normal_FORM}).
Finally, since the transformation of Eq.~(\ref{Local_4proof}) is local and
symplectic, all the determinants relative to the blocks and the global
matrix are preserved, so that Eqs.$~$(\ref{local_invariants}) and$~$(\ref%
{Det_invariant}) are trivially implied.$~\blacksquare $

\section{Genuineness of a two-mode correlation matrix\label{SECgenuine}}

By applying the symplectic tools of the previous Sec.~\ref{ToolsSEC}, we can
now derive very simple algebraic conditions for characterizing the
genuineness of a two-mode CM. In other words, starting from a generic $%
4\times 4$\ real and symmetric matrix $[\mathbf{V}\in \mathcal{S}(4,\mathbb{R%
})]$, we give the algebraic conditions that such a matrix must satisfy in
order to represent the CM of two bosonic modes, i.e., a bona fide two-mode
quantum CM $[\mathbf{V}\in q\mathcal{CM}(2)]$. As a consequence of
Williamson's theorem, we have the following algebraic conditions in terms of
\textit{global} symplectic invariants \cite{AlexHeis}.

\begin{theorem}
\label{TheoGLOBAL}An arbitrary $\mathbf{V}\in \mathcal{S}(4,\mathbb{R})$ is
a quantum CM if and only if it satisfies%
\begin{equation}
\mathbf{V}>0~,~\nu _{-}\geq 1~,  \label{Algebraic1}
\end{equation}%
or, equivalently,%
\begin{equation}
\mathbf{V}>0~,~\det \mathbf{V}\geq 1~,~\Delta (\mathbf{V})\leq 1+\det
\mathbf{V}~.  \label{Algebraic2}
\end{equation}
\end{theorem}

\noindent \textbf{Proof.~}For every $\mathbf{V}\in \mathcal{P}(4,\mathbb{R})$%
, the application of Williamson's theorem to Eq.~(\ref{BONA_FIDE}) gives $%
\mathbf{V}+i\mathbf{\Omega }\geq 0\Longleftrightarrow \nu _{-}\geq 1$
(recalling that $\nu _{-}$ is the \emph{smallest} symplectic eigenvalue).
Since $\mathbf{V}+i\mathbf{\Omega }\geq 0\Longrightarrow \mathbf{V}>0$ (see
Lemma 1), we can write $\mathbf{V}+i\mathbf{\Omega }\geq
0\Longleftrightarrow (\mathbf{V}>0~\wedge ~\nu _{-}\geq 1)$ which proves the
bona fide condition of Eq.~(\ref{Algebraic1}) for a generic $\mathbf{V}\in
\mathcal{S}(4,\mathbb{R})$. Under the definite positivity assumption $%
\mathbf{V}>0$, one can also use Eq.~(\ref{symple}) to prove the equivalences%
\begin{gather}
\nu _{-}\geq 1\Longleftrightarrow \Delta (\mathbf{V})-2\geq \sqrt{\Delta (%
\mathbf{V})^{2}-4\det \mathbf{V}}  \notag \\
\Longleftrightarrow \max \{2,2\sqrt{\det \mathbf{V}}\}\leq \Delta (\mathbf{V}%
)\leq 1+\det \mathbf{V}  \notag \\
\Longleftrightarrow \left\{
\begin{array}{c}
\det \mathbf{V}\geq 1~ \\
2\sqrt{\det \mathbf{V}}\leq \Delta (\mathbf{V})\leq 1+\det \mathbf{V~.}%
\end{array}%
\right.  \label{Conditions1}
\end{gather}%
According to Eq.~(\ref{Global_Inv}) the condition $2\sqrt{\det \mathbf{V}}%
\leq \Delta (\mathbf{V})$ in Eq.~(\ref{Conditions1}) corresponds to $2\nu
_{-}\nu _{+}\leq \nu _{-}^{2}+\nu _{+}^{2}$, which is trivially satisfied.
Then, we have%
\begin{equation}
\nu _{-}\geq 1\Longleftrightarrow \left\{
\begin{array}{c}
\det \mathbf{V}\geq 1~ \\
\Delta (\mathbf{V})\leq 1+\det \mathbf{V~,}%
\end{array}%
\right.
\end{equation}%
which states the equivalence between to Eqs.~(\ref{Algebraic1}) and~(\ref%
{Algebraic2}), where the underlying assumption $\mathbf{V}>0$ is also shown.$%
~\blacksquare $

Notice that, crucially, Williamson's theorem could be applied to $\mathbf{V}$
\emph{because of its definite positivity}, which thus implies the existence
of well-defined symplectic eigenvalues \cite{notpos_note}. The condition $%
\nu _{-}\geq 1$ alone is therefore not, by itself, fully equivalent to the
uncertainty principle $\mathbf{V}+i\mathbf{\Omega }\geq 0$, unless definite
positivity is also assumed. The essential role of the prescription $\mathbf{V%
}>0$, often neglected in the literature, is especially clear in the
formulation of Eq.~(\ref{Algebraic2}): in fact, the other two inequalities
in Eq.~(\ref{Algebraic2}) only depend on the squared symplectic eigenvalues
and cannot thus distinguish between positive and negative eigenvalues.

Besides the algebraic requirements of the previous theorem, we can derive an
alternative set of conditions by applying the reduction to \emph{standard
form}. These new conditions are expressed in terms of \textit{local}
symplectic invariants.

\begin{theorem}
\label{TheoLOCAL}An arbitrary
\begin{equation}
\mathbf{V}=\left(
\begin{array}{cc}
\mathbf{A} & \mathbf{C} \\
\mathbf{C}^{T} & \mathbf{B}%
\end{array}%
\right) \in \mathcal{S}(4,\mathbb{R})
\end{equation}%
is a quantum CM if and only if it satisfies%
\begin{gather}
\mathbf{A},\mathbf{B}>0~,  \label{AeB} \\
\Delta (\mathbf{V})\leq 1+\det \mathbf{V}~,  \label{cond1} \\
2\sqrt{\det \mathbf{A}\det \mathbf{B}}+(\det \mathbf{C})^{2}\leq \det
\mathbf{V}+\det \mathbf{A}\det \mathbf{B~}.  \label{cond2}
\end{gather}
\end{theorem}

\bigskip

\noindent \textbf{Proof.~}Under the assumption $\mathbf{A},\mathbf{B}>0$,
the matrix $\mathbf{V}$ can be reduced to the standard form $\mathbf{V}^{I}$
of Eq.~(\ref{normal_FORM}) via a local symplectic transformation $\mathbf{S}$%
. Since $\mathbf{S\Omega S}^{T}=\mathbf{\Omega }$, the Heisenberg principle
can be written in the equivalent form \cite{Algebra2}%
\begin{equation}
\mathbf{V}+i\mathbf{\Omega }\geq 0\Longleftrightarrow \mathbf{V}^{I}+i%
\mathbf{\Omega }\geq 0~.
\end{equation}%
Since the matrix $\mathbf{V}^{I}+i\mathbf{\Omega }$ is Hermitian, its four
eigenvalues $\lambda _{+}^{+},\lambda _{-}^{+},\lambda _{+}^{-},\lambda
_{-}^{-}$ are real. It is then easy to show that%
\begin{eqnarray}
2\lambda _{\pm }^{+} &=&a+b+\sqrt{\mu \pm 2\sqrt{\nu }}~,  \label{eige1} \\
2\lambda _{\pm }^{-} &=&a+b-\sqrt{\mu \pm 2\sqrt{\nu }}~,  \label{eige2}
\end{eqnarray}%
where%
\begin{equation}
\mu :=4+(a-b)^{2}+2(c_{+}^{2}+c_{-}^{2})\geq 4~,  \label{mu}
\end{equation}%
and
\begin{equation}
\nu :=4(a-b)^{2}+(c_{+}+c_{-})^{2}[4+(c_{+}-c_{-})^{2}]\geq 0~.  \label{ni}
\end{equation}%
Since $\lambda _{+}^{-}$ is the minimum eigenvalue, we have that
\begin{gather}
\mathbf{V}^{I}+i\mathbf{\Omega }\geq 0\Longleftrightarrow \lambda
_{+}^{-}\geq 0\Longleftrightarrow a+b-\sqrt{\mu +2\sqrt{\nu }}\geq 0  \notag
\\
\Longleftrightarrow \left\{
\begin{array}{l}
(a+b)^{2}\geq \mu +2\sqrt{\nu } \\
a+b\geq 0%
\end{array}%
\right. \Longleftrightarrow \left\{
\begin{array}{l}
\lbrack (a+b)^{2}-\mu ]^{2}\geq 4\nu \\
(a+b)^{2}-\mu \geq 0 \\
a+b\geq 0~.%
\end{array}%
\right.  \label{equivalence3}
\end{gather}%
Last condition $a+b\geq 0$ in Eq.~(\ref{equivalence3}) is trivially included
in $\mathbf{A}>0$ and $\mathbf{B}>0$ which gives $a>0$ and $b>0$ (for
congruence with the diagonal matrices $a\mathbf{I}$ and $b\mathbf{I}$). The
other two conditions
\begin{equation}
\lbrack (a+b)^{2}-\mu ]^{2}\geq 4\nu ~,  \label{First_Con}
\end{equation}%
and
\begin{equation}
(a+b)^{2}-\mu \geq 0~,  \label{Sec_Cond}
\end{equation}%
can be recast in terms of the local symplectic invariants. In fact, by
inserting Eqs.~(\ref{mu}) and~(\ref{ni}) in Eq.~(\ref{First_Con}), we get
\begin{equation}
a^{2}+b^{2}+2c_{+}c_{-}\leq (ab-c_{+}^{2})(ab-c_{-}^{2})+1~,  \label{rel1}
\end{equation}%
which is equivalent to Eq.~(\ref{cond1}) by using Eq.~(\ref{Det_invariant})
and $\Delta (\mathbf{V})=\Delta (\mathbf{V}^{I})=a^{2}+b^{2}+2c_{+}c_{-}$.
Finally, by inserting Eq.~(\ref{mu}) in Eq.~(\ref{Sec_Cond}), we get $%
2ab-c_{+}^{2}-c_{-}^{2}\geq 2$ which is equivalent to%
\begin{equation}
2a^{2}b^{2}-ab(c_{+}^{2}+c_{-}^{2})\geq 2ab~,
\end{equation}%
since $ab>0$. In terms of local symplectic invariants, last inequality is
equal to%
\begin{equation}
2\det \mathbf{A}\det \mathbf{B}-I_{4}\geq 2\sqrt{\det \mathbf{A}\det \mathbf{%
B}}~,  \label{Rel_withI4}
\end{equation}%
where
\begin{equation}
I_{4}:=\mathrm{Tr}(\mathbf{A\boldsymbol{\omega }C}\boldsymbol{\omega }%
\mathbf{B\boldsymbol{\omega }C}^{T}\mathbf{\boldsymbol{\omega }}%
)=ab(c_{+}^{2}+c_{-}^{2})  \label{Eq_I4}
\end{equation}%
is another local symplectic invariant. In fact, the quantity $I_{4}$ is
connected to the other local symplectic invariants by

\begin{equation}
\det \mathbf{V}=\det \mathbf{A}\det \mathbf{B}+\left( \det \mathbf{C}\right)
^{2}-I_{4}~,  \label{I4_equation}
\end{equation}%
which holds for every $\mathbf{V}\in \mathcal{S}(4,\mathbb{R})$. Using Eq.~(%
\ref{I4_equation}) in Eq.~(\ref{Rel_withI4}), we then get Eq.~(\ref{cond2}).$%
~\blacksquare $

\section{Separability of a two-mode correlation matrix\label{SECseparability}%
}

Few years ago, Ref.~\cite{Simon} showed how to extend the partial
transposition and the Peres entanglement criterion \cite{Peres} to bipartite
bosonic systems. In fact, partial transposition $\mathrm{PT}:\rho
_{AB}\longrightarrow \tilde{\rho}_{AB}$ corresponds in phase space to a
\textquotedblleft local time reversal\textquotedblright\ which inverts the
momentum of only one of two subsystems. This means that we have the
following transformation for the Wigner function%
\begin{equation}
\mathrm{PT}:W(\mathbf{x})\longrightarrow \tilde{W}(\mathbf{x}):=W(\mathbf{%
\Lambda x})~,  \label{Partial_refle}
\end{equation}%
where%
\begin{equation}
\mathbf{\Lambda }:=\left(
\begin{array}{cc}
1 &  \\
& 1%
\end{array}%
\right) \oplus \left(
\begin{array}{cc}
1 &  \\
& -1%
\end{array}%
\right) ~\mathbf{.}  \label{Lambda}
\end{equation}%
For the corresponding CM $\mathbf{V}\in q\mathcal{CM}(2)$, the PT
transformation is given by%
\begin{equation}
\mathrm{PT}:\mathbf{V}\longrightarrow \mathbf{\tilde{V}:=\Lambda V\Lambda ~},
\label{V_tilde}
\end{equation}%
where the partially transposed matrix $\mathbf{\tilde{V}}$ belongs to $%
\mathcal{P}(4,\mathbb{R})$ \cite{Algebra2}\ but not necessarily to $q%
\mathcal{CM}(2)$. By writing $\mathbf{V}$ in the blockform of Eq.~(\ref%
{CM_2modes}), one easily checks that the action of the PT transformation $%
\mathbf{\Lambda }$ reduces to the following sign flip%
\begin{equation}
\det \mathbf{C\rightarrow }-\det \mathbf{C~,}
\end{equation}%
at the level of the local symplectic invariants. As a consequence, the
positive-definite matrix $\mathbf{\tilde{V}}$ has%
\begin{equation}
\Delta (\mathbf{\tilde{V}})=\det \mathbf{A}+\det \mathbf{B}-2\det \mathbf{C}%
:=\tilde{\Delta}\mathbf{(\mathbf{V})~,}  \label{sera_tilde}
\end{equation}%
and symplectic eigenvalues%
\begin{equation}
\tilde{\nu}_{\pm }=\sqrt{\frac{\tilde{\Delta}(\mathbf{V})\pm \sqrt{\tilde{%
\Delta}(\mathbf{V})^{2}-4\det \mathbf{V}}}{2}~.}
\end{equation}%
Once the PT transformation has been extended, also the Peres criterion can
be consequently extended via the logical implication%
\begin{gather}
\rho _{AB}~\text{separable}\Longrightarrow \tilde{\rho}_{AB}\in \mathcal{D}(%
\mathcal{H})  \notag \\
\Longrightarrow \mathbf{\tilde{V}}\in q\mathcal{CM}(2)~,
\end{gather}%
which becomes an equivalence for Gaussian states under $1\times n$ mode
bipartitions \cite{Simon,wernerwolf}.

\begin{theorem}[Separability]
\label{TheoSEPCM}Let us consider a Gaussian state $\rho _{AB}$ with CM $%
\mathbf{V}\in q\mathcal{CM}(2)$. Then, $\rho _{AB}$ is separable if and only
if%
\begin{equation}
\mathbf{\tilde{V}}\in q\mathcal{CM}(2)~,  \label{separa1}
\end{equation}%
or, equivalently,
\begin{equation}
\tilde{\nu}_{-}\geq 1~,  \label{separa2}
\end{equation}%
or, equivalently,%
\begin{equation}
\tilde{\Delta}\mathbf{(\mathbf{V})}\leq 1+\det \mathbf{V~.}  \label{separa3}
\end{equation}
\end{theorem}

\textbf{Proof.}~The proof of Eq.~(\ref{separa1}) follows exactly the same
steps of the one in Ref.~\cite{Simon}, where the P-representation is
exploited (see Ref.~\cite{Fujikawa} for recent connections between
P-representation and separability.) In order to prove Eqs.~(\ref{separa2})
and~(\ref{separa3}), let us apply Theorem~\ref{TheoGLOBAL} to the
positive-definite matrix $\mathbf{\tilde{V}}\in \mathcal{P}(4,\mathbb{R})$.
Then, we get
\begin{equation}
\mathbf{\tilde{V}}\in q\mathcal{CM}(2)\Longleftrightarrow \tilde{\nu}%
_{-}\geq 1\Longleftrightarrow \left\{
\begin{array}{c}
\det \mathbf{\tilde{V}\geq }1 \\
\tilde{\Delta}(\mathbf{V})\leq 1+\det \mathbf{\tilde{V}}%
\end{array}%
\right.  \label{Equiv_proof}
\end{equation}%
where Eq.~(\ref{separa2}) is trivially proven. Now, since $\det \mathbf{%
\tilde{V}=}\det (\mathbf{\Lambda V\Lambda })=\det \mathbf{V}\geq 1$, the
first condition in Eq.~(\ref{Equiv_proof}) is always satisfied and,
therefore, the separability condition is reduced to Eq.~(\ref{separa3}).~$%
\blacksquare $

Let us now derive the algebraic conditions that a generic symmetric matrix
must satisfy to represent the CM of a separable or entangled Gaussian state.
The following corollary gives an easy recipe to check if a symmetric matrix
is a good or bad candidate for this aim.

\begin{corollary}
\label{CORO_Long}An arbitrary
\begin{equation}
\mathbf{V}=\left(
\begin{array}{cc}
\mathbf{A} & \mathbf{C} \\
\mathbf{C}^{T} & \mathbf{B}%
\end{array}%
\right) \in \mathcal{S}(4,\mathbb{R})
\end{equation}%
represents the CM of a separable Gaussian state if and only if it satisfies%
\begin{equation}
\mathbf{V}>0~,~\nu _{-}\geq 1~,~\tilde{\nu}_{-}\geq 1~,  \label{symSEP1}
\end{equation}%
or, equivalently,%
\begin{equation}
\mathbf{V}>0~,~\det \mathbf{V}\geq 1~,~\Gamma (\mathbf{V})\leq 1+\det
\mathbf{V~,}  \label{symSEP2}
\end{equation}%
or, equivalently,%
\begin{gather}
\mathbf{A},\mathbf{B}>0~,~\Gamma (\mathbf{V})\leq 1+\det \mathbf{V~,}
\label{symSEP3} \\
2\sqrt{\det \mathbf{A}\det \mathbf{B}}+(\det \mathbf{C})^{2}\leq \det
\mathbf{V}+\det \mathbf{A}\det \mathbf{B~,}  \label{symSEPbis}
\end{gather}%
where $\Gamma (\mathbf{V}):=\det \mathbf{A}+\det \mathbf{B}+2\left\vert \det
\mathbf{C}\right\vert $. Instead, it represents the CM of an entangled
Gaussian state if and only if it satisfies%
\begin{equation}
\mathbf{V}>0~,~\nu _{-}\geq 1~,~\tilde{\nu}_{-}<1~,  \label{symENT1}
\end{equation}%
or, equivalently,%
\begin{equation}
\mathbf{V}>0~,~\det \mathbf{V}\geq 1~,~\Delta (\mathbf{V})\leq 1+\det
\mathbf{V}<\tilde{\Delta}(\mathbf{V})\mathbf{~,}  \label{symENT2}
\end{equation}%
or, equivalently,%
\begin{gather}
\mathbf{A},\mathbf{B}>0~,~\Delta (\mathbf{V})\leq 1+\det \mathbf{V}<\tilde{%
\Delta}(\mathbf{V})~,  \label{symENT3} \\
2\sqrt{\det \mathbf{A}\det \mathbf{B}}+(\det \mathbf{C})^{2}\leq \det
\mathbf{V}+\det \mathbf{A}\det \mathbf{B~.}  \label{symENT3bis}
\end{gather}
\end{corollary}

\textbf{Proof.}~In order to represent the CM\ of a separable Gaussian state,
the symmetric matrix $\mathbf{V}\in \mathcal{S}(4,\mathbb{R})$ must
simultaneously satisfy
\begin{equation}
\mathbf{V}\in q\mathcal{CM}(2)~,~\mathbf{\tilde{V}}\in q\mathcal{CM}(2)~.
\end{equation}%
The bona fide condition $\mathbf{V}\in q\mathcal{CM}(2)$ is equivalently
expressed by the conditions of Eqs.~(\ref{Algebraic1}) and~(\ref{Algebraic2}%
) in Theorem~\ref{TheoGLOBAL}. Then, for every $\mathbf{V}\in q\mathcal{CM}%
(2)$, the separability condition $\mathbf{\tilde{V}}\in q\mathcal{CM}(2)$ is
equivalent to Eqs.~(\ref{separa2}) and~(\ref{separa3}) in Theorem~\ref%
{TheoSEPCM}. By combining Eq.~(\ref{Algebraic1}) with Eq.~(\ref{separa2}),
and Eq.~(\ref{Algebraic2}) with Eq.~(\ref{separa3}), one easily gets Eqs.~(%
\ref{symSEP1}) and~(\ref{symSEP2}), where $\max \{\Delta (\mathbf{V}),\tilde{%
\Delta}(\mathbf{V})\}\leq 1+\det \mathbf{V}\Longleftrightarrow \Gamma (%
\mathbf{V})\leq 1+\det \mathbf{V}$. According to Theorem~\ref{TheoSEPCM},
for every $\mathbf{V}\in q\mathcal{CM}(2)$ the entanglement condition is
expressed by $\tilde{\nu}_{-}<1$ or, equivalently, by $\tilde{\Delta}\mathbf{%
(\mathbf{V})}>1+\det \mathbf{V}$. Then, it is trivial to derive the
corresponding Eqs.~(\ref{symENT1}) and~(\ref{symENT2}). The proof of Eqs.~(%
\ref{symSEP3}-\ref{symSEPbis}) and~(\ref{symENT3}-\ref{symENT3bis}) is the
same as before except that now we have to combine the Eqs.~(\ref{AeB}), (\ref%
{cond1}) and~(\ref{cond2}) of Theorem~\ref{TheoLOCAL} with Eq.~(\ref{separa3}%
) for the separability and with $\tilde{\Delta}\mathbf{(\mathbf{V})}>1+\det
\mathbf{V}$ for the entanglement.$~\blacksquare $

\section{Relation with the previous work by Simon\label{Comparison}}

In order to make a direct comparison with the previous work by Simon~\cite%
{Simon} , we have to specify some of our results, given for arbitrary
symmetric matrices $\mathbf{V}\in \mathcal{S}(4,\mathbb{R})$, to the case of
positive-definite matrices, i.e., $\mathbf{V}\in \mathcal{P}(4,\mathbb{R})$.
As an immediate consequence of Theorem~\ref{TheoGLOBAL}, we have the
following result.

\begin{corollary}
\label{CORO_global}An arbitrary $\mathbf{V}\in \mathcal{P}(4,\mathbb{R})$ is
a two-mode quantum CM $\mathbf{V}\in q\mathcal{CM}(2)$, i.e., $\mathbf{V}+i%
\mathbf{\Omega }\geq 0$, if and only if
\begin{equation}
\nu _{-}\geq 1~,  \label{EqCor1}
\end{equation}%
or, equivalently,%
\begin{equation}
\det \mathbf{V}\geq 1~,~\Delta (\mathbf{V})\leq 1+\det \mathbf{V}~,
\label{EqCor2}
\end{equation}%
or, equivalently,%
\begin{gather}
\det \mathbf{V}\geq 1~,  \label{EqCor3} \\
\det \mathbf{A}\det \mathbf{B}+\left( 1-\det \mathbf{C}\right)
^{2}-I_{4}\geq \det \mathbf{A}+\det \mathbf{B}~,  \label{EqCor4}
\end{gather}%
where $I_{4}:=\mathrm{Tr}(\mathbf{A\boldsymbol{\omega }C}\boldsymbol{\omega }%
\mathbf{B\boldsymbol{\omega }C}^{T}\mathbf{\boldsymbol{\omega }})$.
\end{corollary}

\bigskip

\noindent \textbf{Proof.~}By applying Theorem~\ref{TheoGLOBAL} under the
assumption $\mathbf{V}>0$, one trivially derives the equivalent conditions
in Eqs.~(\ref{EqCor1}) and~(\ref{EqCor2}). In order to prove Eqs.~(\ref%
{EqCor3}) and~(\ref{EqCor4}), let us reduce the positive-definite matrix $%
\mathbf{V}$ to its standard form of Eq.~(\ref{normal_FORM}). Under local
symplectic transformations, we then have the equivalence%
\begin{eqnarray}
\Delta (\mathbf{V}) &\leq &1+\det \mathbf{V}\Leftrightarrow  \notag \\
a^{2}+b^{2}+2c_{+}c_{-} &\leq &1+(ab-c_{+}^{2})(ab-c_{-}^{2})~.
\label{EqProof1}
\end{eqnarray}%
Note that Eq.~(\ref{EqProof1}) can be equivalently written as%
\begin{equation}
a^{2}b^{2}+\left( 1-c_{+}c_{-}\right) ^{2}-ab(c_{+}^{2}+c_{-}^{2})\geq
a^{2}+b^{2}~.
\end{equation}%
In terms of local symplectic invariants, last relation can be written as in
Eq.~(\ref{EqCor4}), where Eq.~(\ref{Eq_I4}) has been also used.~$%
\blacksquare $

Notice that Eq.~(\ref{EqCor4}) corresponds to the Eq.~(17) of Ref.~\cite%
{Simon}, up to notation factors \cite{SimNOTATION}. In Ref.~\cite{Simon},
this condition is incorrectly claimed to be equivalent to the Heisenberg
principle $\mathbf{V}+i\mathbf{\Omega }\geq 0$ (this equivalence is claimed
under the positivity contraint $\mathbf{V}>0$, which is a sufficient
condition for the reduction to standard form used in the proof of Ref.~\cite%
{Simon}). In order to have a full equivalence with the Heisenberg principle $%
\mathbf{V}+i\mathbf{\Omega }\geq 0$, the supplementary condition of Eq.~(\ref%
{EqCor3}) is mandatory. It is indeed rather simple to construct a
positive-definite matrix $\mathbf{V}\in \mathcal{P}(4,\mathbb{R})$ which
satisfies Eq.~(\ref{EqCor4}) but violates $\mathbf{V}+i\mathbf{\Omega }\geq
0 $. As an example, let us consider the following real and symmetric matrix

\begin{equation}
\mathbf{V}(x)=\frac{1}{2}\left(
\begin{array}{cccc}
1+4x & 0 & -1+4x & 0 \\
0 & 1+4x & 0 & -4x \\
-1+4x & 0 & 1+4x & 0 \\
0 & -4x & 0 & 1+4x%
\end{array}%
\right) ~,  \label{CMEX}
\end{equation}%
which is positive-definite for every $x>0$. It is easy to verify that the
Hermitian matrix $\mathbf{V}+i\mathbf{\Omega }$ has the following real
eigenvalues%
\begin{eqnarray}
\lambda _{\pm } &=&\frac{1}{4}(1+8x\pm \sqrt{17-16x+64x^{2}})~, \\
\theta _{\pm } &=&\frac{1}{4}(3+8x\pm \sqrt{17-16x+64x^{2}})~.
\end{eqnarray}%
Since $\lambda _{-}$ is the minimal eigenvalue, the Heisenberg principle $%
\mathbf{V}+i\mathbf{\Omega }\geq 0$ is equivalent to $\lambda _{-}\geq 0$,
which gives
\begin{equation}
x\geq 1/2~.  \label{NumHeis}
\end{equation}%
Now, let us explicitly compute Eqs.~(\ref{EqCor3}) and~(\ref{EqCor4}). It is
easy to show that Eq.~(\ref{EqCor3}) is equivalent to%
\begin{equation}
x(8x+1)\geq 1~\Leftrightarrow ~x\geq (\sqrt{33}-1)/16\simeq 0.3~,
\label{SimonMissed}
\end{equation}%
while Eq.~(\ref{EqCor4}) (Simon's genuineness condition) is equivalent to%
\begin{equation}
\frac{1}{2}+x(8x-5)\geq 0~\Leftrightarrow ~0<x\leq \frac{1}{8}~\text{OR}%
~x\geq \frac{1}{2}~.  \label{violationSIMON}
\end{equation}%
From Eq.~(\ref{violationSIMON}), one can see that Simon's condition alone
does not exclude the matrices $\mathbf{V}(x)$ for $0<x\leq 1/8$, which are
clearly unphysical since they violate the Heisenberg condition of Eq.~(\ref%
{NumHeis}). A\ complete equivalence with Eq.~(\ref{NumHeis}) is retrieved by
coupling Eq.~(\ref{violationSIMON}) with Eq.~(\ref{SimonMissed}), where the
latter equation excludes the non-physical region $0<x\leq 1/8$.

This imprecision in Simon's work leads to a common misunderstanding of the
subsequent separability condition [Eq.~(19) of Ref~\cite{Simon}], which in
our notation corresponds to
\begin{equation}
\det \mathbf{A}\det \mathbf{B}+\left( 1-\left\vert \det \mathbf{C}%
\right\vert \right) ^{2}-I_{4}\geq \det \mathbf{A}+\det \mathbf{B}~.
\label{Simon_sepCOND}
\end{equation}%
In this condition, the Heisenberg principle is erroneously claimed to be
included (in fact, it is only partially included). Hence, Simon's
separability condition of Eq.~(\ref{Simon_sepCOND}) is actually valid
\textit{only if }$\mathbf{V}\in q\mathcal{CM}(2)$, i.e., the
positive-definite matrix $\mathbf{V}\in \mathcal{P}(4,\mathbb{R})$ is
already known to be a bona fide quantum CM. In other words, the separability
criterion of Eq.~(\ref{Simon_sepCOND}) must be tested on positive-definite
matrices which are already known to describe the second statistical moments
of a physical quantum state. However, under this assumption of physicality,
Simon's separability criterion of Eq.~(\ref{Simon_sepCOND}) displays a
\textit{redundant} modulus and must be simplified to%
\begin{equation}
\det \mathbf{A}\det \mathbf{B}+\left( 1+\det \mathbf{C}\right)
^{2}-I_{4}\geq \det \mathbf{A}+\det \mathbf{B}~.
\end{equation}

\begin{criterion}[Separability]
Let us consider a two-mode quantum state $\rho _{AB}$ having quantum CM
\begin{equation}
\mathbf{V}=\left(
\begin{array}{cc}
\mathbf{A} & \mathbf{C} \\
\mathbf{C}^{T} & \mathbf{B}%
\end{array}%
\right) \in q\mathcal{CM}(2)~.
\end{equation}%
The separability of $\rho _{AB}$ implies%
\begin{equation}
\tilde{\Delta}(\mathbf{V})\leq 1+\det \mathbf{V~,}  \label{SimonEQUI}
\end{equation}%
or, equivalently,%
\begin{equation}
\det \mathbf{A}\det \mathbf{B}+\left( 1+\det \mathbf{C}\right)
^{2}-I_{4}\geq \det \mathbf{A}+\det \mathbf{B}~.  \label{SimonHolds}
\end{equation}%
In particular, if $\rho _{AB}$ is a Gaussian state, then it is separable if
and only if Eq.~(\ref{SimonEQUI}) [or Eq.~(\ref{SimonHolds})] holds.
\end{criterion}

\bigskip

\noindent \textbf{Proof.~}The proof is straightforward. Since $\mathbf{V}$
is a quantum CM, it is positive-definite and satisfies the condition $\det
\mathbf{V}\geq 1$. Now, suppose that the corresponding two-mode state $\rho
_{AB}$ is separable. Then we have%
\begin{gather}
\rho _{AB}~\text{separable}\Longrightarrow \tilde{\rho}_{AB}\in \mathcal{D}(%
\mathcal{H})\Longrightarrow  \notag \\
\mathbf{\tilde{V}}\in q\mathcal{CM}(2)\Leftrightarrow \mathbf{\tilde{V}}+i%
\mathbf{\Omega }\geq 0~.
\end{gather}%
By applying Corollary~\ref{CORO_global} to the positive-definite matrix $%
\mathbf{\tilde{V}=\Lambda V\Lambda }$, we have%
\begin{equation}
\mathbf{\tilde{V}}+i\mathbf{\Omega }\geq 0\Leftrightarrow \det \mathbf{%
\tilde{V}}\geq 1~,~\Delta (\mathbf{\tilde{V}})\leq 1+\det \mathbf{\tilde{V}~.%
}  \label{Pre_EQ}
\end{equation}%
Since $\det \mathbf{\tilde{V}=}\det \mathbf{V}$, we have that $\det \mathbf{%
\tilde{V}}\geq 1$ is automatically satisfied in the previous Eq.~(\ref%
{Pre_EQ}). Then, we get%
\begin{equation}
\rho _{AB}~\text{separable}\Longrightarrow \tilde{\Delta}(\mathbf{V})\leq
1+\det \mathbf{V~,}
\end{equation}%
where\textbf{\ }$\tilde{\Delta}(\mathbf{V}):=\Delta (\mathbf{\tilde{V}})$ is
defined in Eq.~(\ref{sera_tilde}). Using Eq.~(\ref{I4_equation}), one easily
proves the equivalence between Eqs.~(\ref{SimonEQUI}) and~(\ref{SimonHolds}%
). Finally, the full equivalence which holds for Gaussian $\rho _{AB}$ is a
direct application of Theorem~\ref{TheoSEPCM}.~$\blacksquare $

This criterion represents a simplification of Simon's separability
criterion. Now, it is important to notice that the separability criterion
becomes a bit more involved when arbitrary positive-definite matrices $%
\mathbf{V}\in \mathcal{P}(4,\mathbb{R})$ are considered, without any other a
priori assumption. For a generic $\mathbf{V}\in \mathcal{P}(4,\mathbb{R})$,
both the Heisenberg principle ($\mathbf{V}+i\mathbf{\Omega }\geq 0$) and the
separability property ($\mathbf{\tilde{V}}+i\mathbf{\Omega }\geq 0$) must be
explicitly considered and combined together, in order to get a complete set
of algebraic conditions. Thanks to these conditions, one easily checks when
a positive-definite matrix $\mathbf{V}\in \mathcal{P}(4,\mathbb{R})$ can
represent the quantum CM of a Gaussian state $\rho _{AB}$ which is separable
or entangled. By applying Corollary~\ref{CORO_Long}, we get the following
criterion for positive-definite matrices.

\begin{criterion}
An arbitrary $\mathbf{V}\in \mathcal{P}(4,\mathbb{R})$ represents the CM of
a separable Gaussian state if and only if%
\begin{gather}
\det \mathbf{V}\geq 1~, \\
\Gamma (\mathbf{V})\leq 1+\det \mathbf{V~,}  \label{Nec0}
\end{gather}%
or, equivalently,%
\begin{gather}
\det \mathbf{V}\geq 1~,  \label{Nec1} \\
\det \mathbf{A}\det \mathbf{B}+\left( 1-\left\vert \det \mathbf{C}%
\right\vert \right) ^{2}-I_{4}\geq \det \mathbf{A}+\det \mathbf{B}~.
\label{Nec2}
\end{gather}%
Instead, it represents the CM of an entangled Gaussian state if and only if%
\begin{gather}
\det \mathbf{V}\geq 1~, \\
\Delta (\mathbf{V})\leq 1+\det \mathbf{V}<\tilde{\Delta}(\mathbf{V})~,
\label{Nec3}
\end{gather}%
or, equivalently,%
\begin{gather}
\det \mathbf{V}\geq 1~, \\
\left( 1+\det \mathbf{C}\right) ^{2}<\det \mathbf{A}+\det \mathbf{B-\det
\mathbf{A}\det \mathbf{B}}+I_{4}  \notag \\
\leq \left( 1-\det \mathbf{C}\right) ^{2}.  \label{Eq_forDETC}
\end{gather}
\end{criterion}

The proof is a trivial application of Corollary~\ref{CORO_Long}, together
with Eq.~(\ref{I4_equation}), used to state the equivalences between Eq.~(%
\ref{Nec0}) and Eq.~(\ref{Nec2}), and between Eq.~(\ref{Nec3}) and Eq.~(\ref%
{Eq_forDETC}). According to Eq.~(\ref{Eq_forDETC}), positive-definite
matrices with $\det \mathbf{C}\geq 0$ can only be associated to separable
Gaussian states \cite{Simon}. Notice that the original Simon's separability
criterion, i.e., Eq.~(\ref{Nec2}), must be coupled with the mandatory
condition of Eq.~(\ref{Nec1}) in order to investigate correctly the
separability properties of a generic positive-definite matrix.

\section{Summary\label{SECsummary}}

In Theorem~\ref{TheoGLOBAL}, we have re-derived and explicitly stated all
the precise algebraic conditions a symmetric matrix must satisfy to
represent the CM of a two-mode bosonic (or canonical) quantum system,
including the (critical and often neglected) definite positivity condition.
Such conditions are expressed in terms of global symplectic invariants. In
Theorem~\ref{TheoLOCAL}, we have derived a new and alternative set of
conditions, which are expressed in terms of local symplectic invariants. In
these local conditions the positivity check is restricted to the submatrices
$\mathbf{A}$ and $\mathbf{B}$. Finally, in Theorem~\ref{TheoSEPCM}, the
necessary and sufficient condition for the separability of two-mode Gaussian
states has been reviewed and cast in a compact form. We should stress that
such a condition is valid only under the assumption that physicality is also
met $[\mathbf{V}\in q\mathcal{CM}(2)]$. In Corollary~\ref{CORO_Long}, both
the physicality and separability have been explicitly taken into account.
Then, we have derived a complete set of (global or local)\ conditions that a
generic symmetric matrix must satisfy in order to represent the CM\ of a
separable (or entangled) Gaussian state of two bosonic modes. In Section~\ref%
{Comparison}, some of our results have been specified for positive-definite
matrices and a comparison with the previous results by Simon has been
thoroughly presented.

The rigourous agreement with \emph{all} the conditions here considered
should constitute a constant reference in both the theoretical practice and
the analysis of experimental data involving quantum systems of two canonical
degrees of freedom.

\section{Acknowledgements}

S.P. was supported by a Marie Curie Fellowship of the European Community.
S.L. was supported by the W.M. Keck foundation center for extreme quantum
information theory (xQIT).

\appendix

\section{Simple proof of Williamson's theorem\label{WILLIAMSONTHEO}}

Let us construct the diagonalizing symplectic according to the decomposition

\begin{equation}
\mathbf{S=W}^{1/2}\mathbf{RV}^{-1/2}~,  \label{decomp}
\end{equation}%
with a \emph{suitable} $\mathbf{R}\in \mathcal{SO}(2n)$. In fact%
\begin{gather}
\mathbf{SV\mathbf{S}}^{T}=(\mathbf{W}^{1/2}\mathbf{RV}^{-1/2})~\mathbf{V~(V}%
^{-1/2}\mathbf{R}^{T}\mathbf{W}^{1/2})  \notag \\
=\mathbf{W}^{1/2}\mathbf{R~I~R}^{T}\mathbf{W}^{1/2}=\mathbf{W}^{1/2}\mathbf{%
~I~W}^{1/2}=\mathbf{W~.}
\end{gather}%
Notice that Eq.~(\ref{decomp}) is well-defined since $\mathbf{V}$ and $%
\mathbf{W}$ are positive-definite (therefore, non-singular). However, the
rotation $\mathbf{R}$ in Eq.~(\ref{decomp}) is not arbitrary but must be
chosen in order to make $\mathbf{S}$ symplectic.

Let us apply Eq.~(\ref{decomp}) to the symplectic condition $\mathbf{S\Omega
S}^{T}=\mathbf{\Omega }$. Then, we have%
\begin{gather}
(\mathbf{W}^{1/2}\mathbf{RV}^{-1/2})~\mathbf{\Omega ~(V}^{-1/2}\mathbf{R}^{T}%
\mathbf{W}^{1/2})=\mathbf{\Omega \Leftrightarrow }  \notag \\
\mathbf{\Leftrightarrow R~(V}^{-1/2}\mathbf{\Omega V}^{-1/2})~\mathbf{R}^{T}=%
\mathbf{W}^{-1/2}\mathbf{\Omega W}^{-1/2}\mathbf{\Leftrightarrow }  \notag \\
\mathbf{\Leftrightarrow RXR}^{T}=\mathbf{Y~,}  \label{RXY}
\end{gather}%
where%
\begin{equation}
\mathbf{X:=V}^{-1/2}\mathbf{\Omega V}^{-1/2}~,~\mathbf{Y:=W}^{-1/2}\mathbf{%
\Omega W}^{-1/2}  \label{X_Y}
\end{equation}%
are \emph{antisymmetric} (because $\mathbf{V}$ and $\mathbf{W}$ are
symmetric, while $\mathbf{\Omega }$ is antisymmetric). In particular, we have%
\begin{equation}
\mathbf{Y=}\bigoplus\limits_{k=1}^{n}\left(
\begin{array}{cc}
0 & \nu _{k}^{-1} \\
-\nu _{k}^{-1} & 0%
\end{array}%
\right) ~.  \label{Y_blockDiag}
\end{equation}%
Now the existence of $\mathbf{R}$ in Eq.~(\ref{RXY}) is assured by the
following theorem on the block-diagonalization of real antisymmetric
matrices (specialized to even dimensions) \cite{Horn}

\begin{theorem}
\label{TheoBLOCK}For every $\mathbf{A}=-\mathbf{A}^{T}\in \mathcal{M}(2n,%
\mathbb{R})$, there exists a (unique) $\mathbf{O}\in \mathcal{SO}(2n)$ such
that
\begin{equation}
\mathbf{OAO}^{T}=\bigoplus\limits_{k=1}^{n}a_{k}\boldsymbol{\omega }:=%
\mathbf{\tilde{A}}~,  \label{antiDIAG}
\end{equation}%
where the (unique) block diagonal form $\mathbf{\tilde{A}}$ has $a_{k}>0$.
\end{theorem}

\section{Finding the diagonalizing symplectic matrix\label{FINDING}}

Let us show a possible procedure for deriving the proper rotation $\mathbf{O}
$ that block-diagonalizes a generic antisymmetric matrix $\mathbf{A}$ as in
Theorem~\ref{TheoBLOCK}. We can easily prove the following connection
between the block-diagonalization of $\mathbf{A}$ and its unitary
diagonalization

\begin{theorem}
\label{CorollaryBLOCK}The proper rotation $\mathbf{O}$ performing the
block-diagonalization of Eq.~(\ref{antiDIAG}) is given by%
\begin{equation}
\mathbf{O}=\boldsymbol{\Gamma }\mathbf{U}^{\dagger }~,  \label{OgammaU}
\end{equation}%
where%
\begin{equation}
\boldsymbol{\Gamma }=\frac{1}{\sqrt{2}}\bigoplus\limits_{k=1}^{n}\boldsymbol{%
\gamma }~,~\boldsymbol{\gamma }:=\frac{1}{\sqrt{2}}\left(
\begin{array}{cc}
i & -i \\
1 & 1%
\end{array}%
\right) ~,  \label{GammaBlocks}
\end{equation}%
and $\mathbf{U}$\ is an arbitrary unitary performing the diagonalization of $%
\mathbf{A}$, i.e.,
\begin{equation}
\mathbf{U}^{\dagger }\mathbf{AU}=\bigoplus\limits_{k=1}^{n}ia_{k}\left(
\begin{array}{cc}
-1 &  \\
& 1%
\end{array}%
\right) :=\mathbf{A}_{D}~.  \label{UAU}
\end{equation}
\end{theorem}

\bigskip

\textbf{Proof.~}~First, let us prove how $\mathbf{A}$ can be transformed
into the diagonal form $\mathbf{A}_{D}$ of Eq.~(\ref{UAU}) by a unitary
matrix. From Eq.~(\ref{GammaBlocks}), we have that%
\begin{equation}
\boldsymbol{\gamma }^{\dagger }\mathbf{~}(a_{k}\mathbf{\boldsymbol{\omega }})%
\mathbf{~}\boldsymbol{\gamma }=ia_{k}\left(
\begin{array}{cc}
-1 &  \\
& 1%
\end{array}%
\right) ~.
\end{equation}%
As a consequence, by applying $\boldsymbol{\Gamma }$ to Eq.~(\ref{antiDIAG}%
), we get%
\begin{equation}
\boldsymbol{\Gamma }^{\dagger }\mathbf{OAO}^{T}\boldsymbol{\Gamma }=%
\boldsymbol{\Gamma }^{\dagger }\mathbf{\tilde{A}}\boldsymbol{\Gamma }%
=\bigoplus\limits_{k=1}^{n}ia_{k}\left(
\begin{array}{cc}
-1 &  \\
& 1%
\end{array}%
\right) =\mathbf{A}_{D}~.  \label{GammaA}
\end{equation}%
In other words, there exists a unitary $\mathbf{O}^{T}\boldsymbol{\Gamma }$
that diagonalizes $\mathbf{A}$ according to Eq.~(\ref{UAU}).

Then, let us prove that, for every unitary $\mathbf{U}$ diagonalizing $%
\mathbf{A}$ according to Eq.~(\ref{UAU}), we can write Eq.~(\ref{OgammaU})
where $\mathbf{O}$ performs the block-diagonalization of Eq.~(\ref{antiDIAG}%
). For proving this, let us consider the orthonormal eigenvectors $\{\mathbf{%
u}_{1},\cdots ,\mathbf{u}_{2n}\}$ of $\mathbf{A}$%
\begin{gather}
\mathbf{Au}_{1}=-ia_{1}\mathbf{u}_{1}~,~\mathbf{Au}_{2}=ia_{1}\mathbf{u}%
_{2}~, \\
\vdots  \notag \\
\mathbf{Au}_{2n-1}=-ia_{n}\mathbf{u}_{2n-1}~,~\mathbf{Au}_{2n}=+ia_{n}%
\mathbf{u}_{2n}~,
\end{gather}%
more compactly denoted by $\{\mathbf{u}_{2k-1},\mathbf{u}_{2k}\}_{k=1}^{n}$
with%
\begin{equation}
\mathbf{Au}_{2k-1}=-ia_{k}\mathbf{u}_{2k-1}~,~\mathbf{Au}_{2k}=ia_{k}\mathbf{%
u}_{2k}~.  \label{Eigenvectors}
\end{equation}%
These vectors are unique up to phase factors $\boldsymbol{\varphi }%
:=\{\varphi _{1},\cdots ,\varphi _{2n}\}$, i.e., up to the replacements
\begin{equation}
\mathbf{u}_{2k-1}\rightarrow \mathbf{u}_{2k-1}e^{i\varphi _{2k-1}}~,~\mathbf{%
u}_{2k}\rightarrow \mathbf{u}_{2k}e^{i\varphi _{2k}}~.
\end{equation}%
This means that, for every choice of $\boldsymbol{\varphi }$, we have an
equivalent unitary matrix $\mathbf{U}=\mathbf{U}(\boldsymbol{\varphi })$ in
the diagonalization of $\mathbf{A}$. Now, by conjugating Eq.~(\ref%
{Eigenvectors}), one easily checks that
\begin{equation}
\mathbf{u}_{2k-1}=\mathbf{u}_{2k}^{\ast }~.  \label{u_conjugate}
\end{equation}%
As a consequence, the most general unitary matrix that diagonalizes $\mathbf{%
A}$ has the specific form%
\begin{equation}
\mathbf{U}=\left(
\begin{array}{ccccc}
\mathbf{u}_{2}^{\ast } & \mathbf{u}_{2} & \cdots & \mathbf{u}_{2n}^{\ast } &
\mathbf{u}_{2n}%
\end{array}%
\right) ~.
\end{equation}%
Let us explicitly compute the matrix product $\boldsymbol{\Gamma }\mathbf{U}%
^{\dagger }$. By applying
\begin{equation}
\boldsymbol{\Gamma }=\frac{1}{\sqrt{2}}\left(
\begin{array}{ccc}
\begin{array}{cc}
i & -i \\
1 & 1%
\end{array}
&  & \mathbf{0} \\
& \ddots &  \\
\mathbf{0} &  &
\begin{array}{cc}
i & -i \\
1 & 1%
\end{array}%
\end{array}%
\right)
\end{equation}%
to the conjugate matrix%
\begin{gather}
\mathbf{U}^{\dagger }=\left(
\begin{array}{c}
\mathbf{u}_{2}^{T} \\
\mathbf{u}_{2}^{\dagger } \\
\vdots \\
\mathbf{u}_{2n}^{T} \\
\mathbf{u}_{2n}^{\dagger }%
\end{array}%
\right)  \notag \\
=\left(
\begin{array}{ccccc}
u_{2,1} & u_{2,2} &  & u_{2,2n-1} & u_{2,2n} \\
u_{2,1}^{\ast } & u_{2,2}^{\ast } &  & u_{2,2n-1}^{\ast } & u_{2,2n}^{\ast }
\\
&  & \ddots &  &  \\
u_{2n,1} & u_{2n,2} &  & u_{2n,2n-1} & u_{2n,2n} \\
u_{2n,1}^{\ast } & u_{2n,2}^{\ast } &  & u_{2n,2n-1}^{\ast } &
u_{2n,2n}^{\ast }%
\end{array}%
\right) ~,
\end{gather}%
one explicitly gets%
\begin{equation}
\boldsymbol{\Gamma }\mathbf{U}^{\dagger }=\frac{1}{\sqrt{2}}\left(
\begin{array}{ccccc}
\alpha _{1,1} & \alpha _{1,2} &  & \alpha _{1,2n-1} & \alpha _{1,2n} \\
\beta _{1,1} & \beta _{1,2} &  & \beta _{1,2n-1} & \beta _{1,2n} \\
&  & \ddots &  &  \\
\alpha _{n,1} & \alpha _{n,2} &  & \alpha _{n,2n-1} & \alpha _{n,2n} \\
\beta _{n,1} & \beta _{n,2} &  & \beta _{n,2n-1} & \beta _{n,2n}%
\end{array}%
\right) ~,  \label{GammaU}
\end{equation}%
where%
\begin{equation}
\alpha _{k,j}:=-2\text{\textrm{Im}}(u_{2k,j})~,~\beta _{k,j}:=2\text{\textrm{%
Re}}(u_{2k,j})~.  \label{ImRe}
\end{equation}%
From Eqs.~(\ref{GammaU}) and~(\ref{ImRe}), we have that $\boldsymbol{\Gamma }%
\mathbf{U}^{\dagger }$ is \emph{real} for every choice of $\mathbf{U}$ in
Eq.~(\ref{UAU}), i.e., for every choice of the phases $\boldsymbol{\varphi }$
in the corresponding eigenvectors. More strongly, we have $\boldsymbol{%
\Gamma }\mathbf{U}^{\dagger }\in \mathcal{SO}(2n)$ (since $\boldsymbol{%
\Gamma }\mathbf{U}^{\dagger }$ real implies $\boldsymbol{\Gamma }\mathbf{U}%
^{\dagger }$ orthogonal with $\det =+1$). Then, from Eq.~(\ref{UAU}), we
easily get%
\begin{equation}
\boldsymbol{\Gamma }\mathbf{U}^{\dagger }\mathbf{AU}\boldsymbol{\Gamma }%
^{\dagger }=\boldsymbol{\Gamma }\mathbf{A}_{D}\boldsymbol{\Gamma }^{\dagger
}=\mathbf{\tilde{A}}~.
\end{equation}%
In conclusion, for every diagonalizing unitary $\mathbf{U}$, the proper
rotation $\boldsymbol{\Gamma }\mathbf{U}^{\dagger }$ corresponds to the
\emph{unique} proper rotation $\mathbf{O}$ that performs the
block-diagonalization of Eq.~(\ref{antiDIAG}).$~\blacksquare $

Both Theorem~\ref{TheoBLOCK} and Theorem~\ref{CorollaryBLOCK} can be applied
to the Eq.~(\ref{RXY}), by setting $\mathbf{O=R}$, $\mathbf{A=X}$ and $%
\mathbf{\tilde{A}=Y}$. These theorems allow to reduce the computation of the
rotation $\mathbf{R}$\ in Eq.~(\ref{RXY})\ to a unitary diagonalization. In
fact, we have just to find a unitary $\mathbf{U}$ that diagonalizes $\mathbf{%
X}$, i.e.,%
\begin{equation}
\mathbf{U}^{\dagger }\mathbf{XU}=\bigoplus\limits_{k=1}^{n}i\nu
_{k}^{-1}\left(
\begin{array}{cc}
1 &  \\
& -1%
\end{array}%
\right) ~,
\end{equation}%
and then construct%
\begin{equation}
\mathbf{R=\boldsymbol{\Gamma }\mathbf{U}^{\dagger }}~.
\end{equation}%
Once that we have $\mathbf{R}$, we use Eq.~(\ref{decomp}) to get the
symplectic $\mathbf{S}$. Here is the complete\textbf{\ }algorithm:

\begin{enumerate}
\item Find the symplectic spectrum of $\mathbf{V}$, i.e., its Williamson
form $\mathbf{W}$

\item Compute the matrices $\mathbf{W}^{1/2}$ (immediate) and $\mathbf{V}%
^{-1/2}$ (needs orthogonal diagonalization)

\item Construct the matrix $\mathbf{X:=V}^{-1/2}\mathbf{\Omega V}^{-1/2}$

\item Find the eigenvectors of $\mathbf{X}$ and construct the corresponding
unitary $\mathbf{U}$

\item Compute $\mathbf{R=\boldsymbol{\Gamma }\mathbf{U}^{\dagger }}$

\item Compute $\mathbf{S=W}^{1/2}\mathbf{RV}^{-1/2}$.
\end{enumerate}

\noindent By construction, this algorithm reduces the determination of $%
\mathbf{S}$ to \textit{unitary diagonalizations}. Actually, this task can be
achieved via faster methods when the symplectic spectrum is non-degenerate.
In general, the determination of $\mathbf{S}$\ is equivalent to the
construction of a symplectic basis \cite{Arnold}.


\begin{thebibliography}{99}
\bibitem{cm_note} Often, in the theoretical practice, the second moments of
a bosonic or canonical quantum system are grouped together in a matrix
called \textquotedblleft correlation\textquotedblright\ or \textquotedblleft
covariance\textquotedblright\ matrix.
\bibitem{global} It is trivial to say that a global invariant is also a
local invariant.
\bibitem{Spectrum} In fact, the matrix $i\mathbf{\Omega V}$ is Hermitian
and, therefore, diagonalizable by a unitary transformation. Then, by taking
the modulus of its $2n$ real eigenvalues, one gets the $n$ symplectic
eigenvalues of $\mathbf{V}$.
\bibitem{Algebra1} For every $\mathbf{M}\in \mathcal{M}(n,\mathbb{R})$ there
always exists a pair of proper rotations $\mathbf{R}_{1},\mathbf{R}_{2}\in
\mathcal{SO}(n)$ such that $\mathbf{M}=\mathbf{R}_{1}\mathbf{DR}_{2}^{T}$
with $\mathbf{D}$ diagonal and real.
\bibitem{SimNOTATION} Notice that Ref.~\cite{Simon} adopts the commutation
relations $[\hat{x}_{l},\hat{x}_{m}]=i\mathbf{\Omega }_{lm}$, so that the
variance of the vacuum noise is equal to $1/2$. In this notation, our Eq.~(%
\ref{EqCor4}) becomes $\det \mathbf{A}\det \mathbf{B}+[(1/4)-\det \mathbf{C}%
]^{2}-I_{4}\geq (\det \mathbf{A}+\det \mathbf{B})/4$, which is exactly the
Eq.~(17) of Ref.~\cite{Simon}.
\end{thebibliography}
\end{document}